\documentstyle[aps,preprint,epsfig]{revtex}
\begin{document}
\title{Gravitational memory of natural wormholes}
\author{L. A. Anchordoqui, M. L. Trobo, H. Vucetich, and F. Zyserman}
\address{Departamento de F\'{\i}sica, Universidad Nacional de La Plata\\
CC. 67 (1900) La Plata\\
Argentina}
\maketitle
\begin{abstract}
A traversable wormhole solution of general scalar--tensor
field equations is presented. We have shown, after a numerical analysis for 
the behavior of the scalar field of Brans-Dicke theory, that the solution is 
completely singularity--free. Furthermore,  the analysis of more 
general scalar field dependent coupling constants indicates that 
the gravitational memory phenomenon may play an important role in the fate of 
natural wormholes. 

\noindent{\it PACS number(s): 04.20.Gz, 04.50.+h}

\end{abstract}
\newpage

Notwithstanding various possible exceptions, the mandate of the
energy conditions (encoded in the evolution of the expansion scalar
governed by Raychaudhuri's equation) is a popular tenet of belief in
any current research field \cite{hawking-ellis}. The simplest
possible stellar environment that militate against these conjectures
is usually called a ``wormhole throat'' \cite{visser-book}.

Wormhole physics dates back at least as far as 1916 \cite{flamm},
with punctuated revivals of activity following both
the classical work of Einstein--Rosen \cite{einstein-rosen}
and the seminal paper by Wheeler \cite{wheeler}, but,
the concept of traversable wormhole was not put forward until
1988 with the work by Morris--Thorne (MT) \cite{motho}.
Nowadays, the broad phylum
of wormholes can be precisely defined in terms of null geodesic
congruences propagating outward from, and orthogonal to, an extremal
hypersurface of minimal area known as the wormhole throat \cite{hvwh}.
Unlike MT's embedding approach, this new
characterization (that accurately captures the essence
of the local physics which operates near any wormhole throat) is free
from assumptions about symmetries, asymptotic flattness, embedded diagrams
and even time dependence. We hasten to point out however, that
throughout this article we shall use the word ``wormhole'' to specify the
sub-class of traversable entities introduced by MT \cite{motho}, namely,
a compact $S^2$ tunnel connecting two asymptotically flat
spacetimes. In order to expand such a configuration along the radial 
direction, the derivative of the $S^2$ radius with respect to the proper 
coordinate $l$ must be positive near the throat. Since this derivative is 
proportional to the energy momentum tensor on the radially directed null 
geodesic, a large inward pressure is required on the tunnel to sustain the 
wormhole structure, yielding a violation of the null energy condition.

Very recently the consequences of the energy conditions were confronted
with possible values of the Hubble parameter and the gravitational
redshifts of the oldest stars in the galactic halo \cite{visser}. It
was deduced that for the currently favored values of $H_0$, the
strong energy condition should have been violated sometime between
the formation of the oldest stars and the present epoch.
If also the null energy condition could have been violated and the 
universe does admit wormhole topologies in it, it is very likely that 
gravitational microlensing effects of these ``exotic''
objects could be monitored here on Earth \cite{cramer}.
The search for such a
wormhole signature has been already started, but no conclusive evidence
of wormhole topologies has been found neither in the galactic halo nor at
cosmological distances \cite{MNRASwh}.

On the other hand, wormhole formation at a late cosmic time requires
a Lorentzian topology change, something that appears to be more than
problematic to most physicists because it implies causality
violations \cite{hawking,mothoyu}.
However, if wormholes are not created by
astrophysical processes but were threaded in the embryo of the spacetime,
one could expect a cosmological population of these objects without the
uncomfortable predictions of topology change theorems. Building on
this theme, Hochberg and Kephart proposed a subway network of wormholes,
that might have arisen out during the quantum gravity epoch, as
an alternative solution for the cosmological horizon problem \cite{hk}.

Despite no
universal mechanism to generate a relic density of exotic
matter is well established yet (because of our
ignorance of quantum gravity laws), if wormholes are leftovers from
the primeval explosion, it is much more likely that their creation was
accompanied by the extremely large curvature fluctuations generally
believed to have existed during the Plank scale. For this reason,
alternative theories of gravity have become a commonplace
in wormhole physics. The already discussed scenarios are higher
derivatives theories \cite{hochberg},
non symmetric theories \cite{moffat}, Einstein--Gauss--Bonnet
theory \cite{bhawal-kar} and Brans-Dicke theory
\cite{bdwh}, among others.
The latter actually is a special case of a more general group known as
scalar-tensor theories (STTs) \cite{stt}.
In such a class of theories (and also in the hyperextended
version \cite{diego-pipi}) the equation of motion for test particles is
identical to that of general relativity (GR), but the locally measurable value
of the gravitational constant $G$ is a function of a scalar field which is
in turn determined by the trace of the energy momentum tensor of all other
non-gravitational fields. Recently, STTs have regained popularity in
the understanding of the very early universe. On the one hand, it has been
shown that these theories might drive new forms of inflation \cite{la}.
On the other hand, it has been suggested that an epoch dependent
asymptotic boundary condition of the scalar field may have important
implications in astrophysical objects such as black holes \cite{barrow}
or boson stars \cite{diegoetal}: they may retain
memory of the value of the gravitational constant at the time of formation, 
or evolve changing their configuration according to the variation of $G$.
It, therefore, seems natural that in the context of wormhole physics
this gravitational memory may play an important role. With this in mind, 
we shall search for a wormhole solution in general STTs, and after 
working out a theoretical construction of such a system we shall briefly 
discuss on its possible graviational memory \cite{sttwh}. Let us start by 
deriving the generalized field equations.

The fundamental action can be cast in the following form \cite{will},
\begin{equation}
S = \int \frac{\sqrt{-g}}{16 \pi} d^4x \, \left[ \phi R -
\frac{\omega(\phi)}{\phi} g^{\mu\nu} \phi_{,\mu} \phi_{,\nu} + 16
\pi {\cal  L}_{\rm m} \right],
\label{action}
\end{equation}
where $g_{\mu\nu}$ stands for the metric tensor, $g \equiv$ det$g_{\mu\nu}$,
$R$ is the scalar curvature of the spacetime, $\phi$ is the scalar
field, $\omega (\phi)$ is a dimensionless function which calibrates
the coupling between the scalar field and gravity,
and ${\cal L}_{\rm m}$ is the Lagrangian of the matter content of the system.
Varying the action with respect to the dynamical variables
$g_{\mu\nu}$ and $\phi$ yields the field equations
\begin{equation}
R_{\mu\nu} - \frac{1}{2} g_{\mu\nu} R =
\frac{8 \pi
G_{_\infty}}{\phi} T_{\mu\nu} + \frac{\omega(\phi)}{\phi^2}
\left(\phi_{;\mu} \phi_{;\nu} - \frac{1}{2} g_{\mu\nu}
\phi^{;\alpha} \phi_{;\alpha} \right) +
\frac{1}{\phi} (\phi_{;\mu\nu} - g_{\mu\nu} \Box \phi),
\label{1}
\end{equation}
\begin{equation}
\Box \phi \equiv \phi^{;\alpha}\,\!\!_{;\alpha} = \frac{1}{2 \omega(\phi) + 3}
\left[ 8 \pi G_{_\infty} T - \frac{d\omega}{d\phi} \phi^{;\alpha}
\phi_{;\alpha}  \right],
\label{2}
\end{equation}
where the Ricci tensor reads  
\begin{equation}
R_{\mu\nu}\equiv
\Gamma_{\mu\nu,\alpha}^\alpha - \Gamma_{\mu\alpha,\nu}^\alpha +
\Gamma_{\mu\nu}^\alpha \Gamma_{\alpha\beta}^\beta -
\Gamma_{\mu\beta}^\alpha \Gamma_{\alpha\nu}^\beta,
\label{ricci}
\end{equation}
with the Riemannian conection
\begin{equation}
\Gamma_{\mu\nu}^\alpha = \frac{1}{2} \, g^{\alpha\beta} (g_{\beta\mu,\nu} +
g_{\beta \nu,\mu} - g_{\mu\nu,\beta}).
\end{equation}

We have introduced $T_{\mu\nu}$ as the energy momentum tensor of
matter and all nongravitational fields, and $T$ as its trace. From
now on we assume $T=0$ and we use ``geometrized units''
in which the speed of light and $G_{_\infty}^{^{\rm today}}$ are equal to 
unity. Commas denote partial derivatives with 
respect to the coordinates $x^\mu$,
semicolon mean covariant derivatives, and primes derivatives with respect
to $r$.

The theory can also be represented as Einstein's general relativity,
with a new scalar field as an additional external
non gravitational field that contributes to the curvature of the
spacetime through its energy momentum tensor. This form of the field
equations is derived from (\ref{1}) and (\ref{2})
by means of the conformal transformation $g_{\mu\nu} \equiv
\bar{g}_{\mu\nu} \,\phi_{_\infty}/ \phi$ (in order to keep the next
results as simple as possible we shall assume that $\phi_{_\infty} = 1$, 
and that $\phi$ in the following expressions is dimensionless). With the
transformed metric $\bar{g}_{\mu\nu}$ we can define all required
geometrical objects, namely, $\bar{R}_{\mu\nu}$ is defined as in
(\ref{ricci}) using the metric $\bar{g}_{\mu\nu}$ in place of
$g_{\mu\nu}$, or equivalently, we can compute the transformed Ricci
scalar $\bar{R}$ (for details see, for instance, \cite{wald}).
Substitution of the latter into the basic
action (\ref{action}), and the use of $\sqrt{-g} \equiv \phi^{-2} \sqrt{-\bar{g}}$,
yields the action in the conformal representation
\begin{equation}
S = \int \frac{\sqrt{-\bar{g}}}{16\pi}\,d^4x \left[ \bar{R} -
\bar{g}^{\mu\nu} [\omega(\phi) + \,3/2]  \frac{\phi_{,\mu}
\phi_{,\nu}}{\phi^2} + 3 \bar{g}^{\mu\nu} \left( \frac{\phi_{,\mu}}{\phi} \right)_{;\nu}
+ 16 \pi \frac{{\cal L}_{\rm m}}{\phi^2}\right].
\end{equation}
Hereafter we discard the third term in the action since it involves a total
derivative. Redefining the scalar field by
\begin{equation}
\varphi_{,\alpha} =
|\omega(\phi) + \,3/2 |^{1/2} \,\frac{\phi_{,\alpha}}{\phi},
\label{vir}
\end{equation}
one easily verifies that the action finally adopts the form
\begin{equation}
S = \int \frac{\sqrt{-\bar{g}}}{16\pi}\,d^4x \left[ \bar{R} -
 \bar{g}^{\mu\nu} \varphi_{,\mu}
\varphi_{,\nu} \right] + \bar{S}_{\rm m},
\end{equation}
where $\bar{S}_{\rm m}$ represents the conformal action for the matter fields.
The field equations in the transformed frame read then,
\begin{equation}
\bar{\Box} \varphi = 0,
\end{equation}
and
\begin{equation}
\bar{R}_{\mu\nu} - \frac{1}{2} \bar{g}_{\mu\nu} \bar{R} =  8 \pi
\bar{T}_{\mu\nu} +
\left[\varphi_{;\mu} \varphi_{;\nu} - \frac{1}{2} \bar{g}_{\mu\nu}
\bar{g}^{\alpha\beta} \varphi_{;\alpha} \varphi_{;\beta} \right].
\end{equation}

Let us introduce now the standard form for the
metric tensor of a static spherically symmetric wormhole \cite{motho}
\begin{equation}
ds^2 = - e^{2\Phi(r)} dt^2 + (1-b(r)/r)^{-1} dr^2 + r^2 d\Omega^2,
\end{equation}
where $d\Omega^2$ stands for the  line element of the sphere $S^2$. It
contains
two unknown functions: i) the redshift function  which is
assumed to be such that $e^{2\Phi(r)}$ is never zero
--this ensures the absence of an event horizon-- and ii) the shape
function $b(r)$.

Our choice
for the redshift is fairly simple, $\Phi(r) = -\alpha/r$,
with $\alpha$ a positive constant. Now, if $\bar{T}_\mu^\nu = $
diag [$\bar{\rho}, -\bar{\tau}, \bar{p}, \bar{p}$] (energy density,
tension and pressure respectively), the nonvanishing components
of the fields equations are
\begin{equation}
8 \pi \bar{\rho} = - \frac{\alpha^2}{r^4} \left( 1 - \frac{b(r)}{r} \right) +
\frac{\alpha}{2} \frac{b'(r)}{r^3} - \frac{\alpha}{2} \frac{b(r)}{r^4},
\end{equation}
\begin{equation}
8 \pi \bar{\tau} = \left( 1 - \frac{b(r)}{r} \right)
\left( -\frac{2 \alpha}{r^3}
+ \frac{\alpha^2}{r^4} + (\varphi')^2 \right) - \frac{\alpha}{2}
\frac{b'(r)}{r^3} + \frac{\alpha}{2} \frac{b(r)}{r^4} -
\frac{b'(r)}{r^2} + \frac{b(r)}{r^3},
\end{equation}
\begin{equation}
8 \pi \bar{p} = \frac{b'(r)}{2r^2} - \frac{\alpha}{r^3} \left( 1 -
\frac{b(r)}{r}\right) + \frac{b(r)}{2r^3},
\end{equation}
\begin{equation}
(\varphi')^2 = \kappa^2 \, \frac{e^{2\alpha/r}}{r^3 [r - b(r)]},
\label{fabio}
\end{equation}
with $\kappa$ an integration constant. Imposing the traceless
constraint we first obtain
\begin{equation}
b'(r) + b(r) \frac{2\alpha^2 - \alpha r}{r^2 (\alpha + 2r)} - \frac{2
\alpha^2 +  \kappa^2 e^{2\alpha/r}}{r (\alpha +2r)} = 0.
\label{oeste}
\end{equation}
After the change of variables, $x = 2 + \alpha/r$, Eq. (\ref{oeste})
integrates straightforwardly to
\begin{equation}
b(x) = \alpha \left( \frac{1}{x} + \frac{2}{x^2} + \frac{3}{x^3} + \frac{3}{x^4}
+ \frac{3}{2x^5}  -  \frac{\kappa^2}{5\alpha^2} e^{2 \,(x-2)}
+ \frac{{\cal K}}{x^5} e^{2 \, (x-2)}\right),
\label{shape}
\end{equation}
where ${\cal K}$ is an integration constant to be determined.

Every wormhole (recall that we are dealing with MT's wormhole),
by its definition, must have a minimum radius
$r_{_{\rm th}}$ (the wormhole throat) at which its embedded surface is
vertical, and consequently the radial coordinate is ill 
behaved. However, as usual one defines
the proper radial distance
\begin{equation}
l(r) = \pm \int \frac{dr}{[1-b(r)/r]^{1/2}}
\end{equation}
which is well behaved throughout the spacetime, i.e.,
$\forall r$, $1-b(r)/r \geq
0$. For the space to be asymptotically flat, far from the
throat we require that $b(r)/r\, \rightarrow 0$ as $l  \rightarrow \pm \infty$.
Notice that our shape function satisfies this condition irrespective
of the choice of ${\cal K}$. In addition, the wormhole must flare
outward near the throat. Stated mathematically, $b(r)/r \leq 1$, with
the equality holding at the throat. This indeed constraints the value
of ${\cal K}$ but not the absolute size of the throat which depends
further on the parameter $\alpha$,
\begin{equation}
{\cal K} = e^{-2(x_{_{\rm th}}-2)}
\frac{x^3_{_{\rm th}}+3x^2_{_{\rm th}}+4.5x_{_{\rm th}}+3}{x_{_{\rm
th} }-2} + 
\frac{\kappa^2 x^5_{_{\rm th}}}{5\alpha^2}.
\end{equation}

The aforementioned properties of $b(r)$, together with the condition
on $\Phi$, entail that the metric tensor describes two asymptotically
flat spacetimes joined by a throat. Notice that the solution goes over 
GR if $\varphi$ is equal to a constant \cite{kar-sahdev}.

We turn now to the analysis of the scalar field with two possible 
courses of cosmological evolution as background: i) The matter that
threads the wormhole remains completely static decoupled from the 
cosmological expansion of the universe around. This regime actually cannot be 
completely correct since the asymptotic gravitational constant does evolve. 
However, in a more realistic situation one could just na\"{\i}vely expect
that the effect of $\phi$ at the throat varies much slower than the cosmological 
evolution of $G$. 
ii) The  energy momentum tensor adjusts its parameters to the change of $G$ 
at short intervals of time when comparing with the scale 
of cosmological evolution.

At this stage it is worthwhile to remark that despite some kind of 
``evolution'' is implicit between the matter and the scalar sectors,
we are dealing with a static metric during the whole proccess. This means
that the location of the extremal hypersurfaces which might see any in-coming 
and/or out-going null geodesic always coalesce in the center of the 
wormhole \cite{hvwh}. We are not going to discuss here about the 
``evolution'' of the stress energy tensor nor the stability of the 
wormhole solution, being this beyond the scope of this letter. 
Instead we content ourselves analyzing snapshots of wormhole-like 
spacetimes. Afterwards, assuming that these configurations lay in
the path of the real cosmological evolution, we are able to perform 
a preliminary research on gravitational memory of natural wormholes.

Replacing Eq. (\ref{vir})
in Eq.(\ref{fabio}) we obtain an ordinary differential equation
for $\phi$ and $\omega$,
\begin{equation}
\sqrt{\omega(\phi) + 3/2}\, \frac{\phi'}{\phi} = \kappa \, 
\frac{e^{\alpha/r}}{r^2}\, \frac{1}{[1-b(r)/r]^{1/2}}.
\label{ictp}
\end{equation}
Although there is no {\it a priori} prescription 
about the coupling function, to make a proper study of a varying 
gravitational strength it is imperative to operate within a self--consistent 
scenario. In this direction, cosmological solutions consistent with weak 
field limits were derived in Ref. \cite{barrow-parsons}. This group
(hereafter referred to as type I solutions) allows $\phi$ to evolve into 
asymptotic values consistent with current weak field limits independently 
of the imposed boundary condition, i.e., 
earlier values of $\phi_{_\infty}$ may be equal to, less than, or bigger 
than 1. A second class of coupling functions may be  described by a 
power-law dependence, $\omega(\phi) \propto \phi^n$ (type II solutions)  
\cite{barrow-mimoso}. In this case the parameter space is not 
constrained by weak field tests, but it is by nucleosynthesis processes 
($n\geq4$ is compatible with the observed abundance of primordial 
helium \cite{dft}). In order to solve Eq. (\ref{ictp}) we had recourse 
to an adaptive stepsize fifth order Runge-Kutta method \cite{nr} setting 
$\alpha=r_{_{\rm th}}$=1.

In the case of Brans Dicke theory, solutions were found  independently of the 
value of $\omega$. Some examples are plotted in Fig. 1. As expected the 
solution becomes constant as $\omega \rightarrow \pm \infty$, i.e., 
the theory goes over GR.
The slope of $\phi \rightarrow \infty$ as $r \rightarrow r_{_{\rm th}}$, but
$\phi$ remains positive in all cases. However, within this framework solar 
system observations imply that the coupling constant need to be very large, 
actually, $|\omega|>500$ \cite{will}. Thus, possible modifications to the 
wormhole parameters are negligible whether or not there is gravitational 
memory.

For type I solutions  \cite{barrow-parsons} we found that  $\phi$ remains 
constant throughout the spacetime. Thus, assuming that in any wormhole 
geometry $\phi''(x)<0$ in the entire domain of $x$, to mantain the wormhole
opened (within a hypothetical $G$ time--varying 
scenario) the matter sector must evolve adjusting its value to the ADM 
mass with the changing $G$ (for definition of ADM mass see, for instance, 
chapter 11 of Ref. \cite{visser-book}). 
 
When the coupling function was assumed to be $2 \omega+3 = \omega_0 (\phi/
\phi_{_\infty}^{^{\rm today}})^n$
\cite{barrow-mimoso} it became apparent that the existence of the solution
is strongly correlated with the values of $\omega_0$ and $n$. In some
cases the slope of the scalar field becomes infinite before reaching
the throat (e.g. $\omega_0$ = 1, $n$ = 1), yielding no wormhole solutions.
Nonetheless, well behaved solutions are attainable for bounded regions of 
the parameter space. In particular, we have plotted in Fig. 1 the one 
corresponding to $\omega_0$ = 100, $n$ = 4. 

To give a step further 
we  changed the asymptotic values for the scalar field and repeated the 
previous analysis (notice that because of the change in the boundary 
condition, regions of the parameter space which have shown singular 
solutions might acquire analyticity and vice versa).  
Our results are listed in Table I. As was expected, for type I solutions, if 
one considers the 
final stages of the cosmological evolution,   
the rate of change for the scalar at the throat is bigger than 
its asymptotic behavior. For instance, a change of 
$\phi_{_\infty} \sim 10 \%$ yields a variation of 69 $\%$ at the throat. 
Remarkably, at the very early universe the situation is turned over, 
i.e., a $10 \%$ in the variation of $\phi_{_\infty}$ corresponds to a $5\%$
in the variation of $\phi_{_{\rm th}}$.
For coupling ``constants'' (type II) compatible with nucleosynthesis tests, 
the rate of change of $\phi$ is similar at $x=2$ 
and $x=3$, whereas 
if, $2\omega+3 = 10 \, \phi/\phi_{_\infty}^{^{\rm today}}$,  the variation at 
the wormhole throat may be slower than that at $r \rightarrow \infty$. All 
in all, contrary to type I, for type II solutions one could expect that the 
local value of $G$ within the wormhole might be preserved while its 
asymptotic value evolves with a cosmological rate. It would also be desirable 
a well behaved cosmological evolution of such $\omega (\phi)$. Unfortunately, 
one need to be aware that the last scheme is not consistent with 
nucleosynthesis tests.  In any case, it is straightforward that gravitational 
memory phenomenon -- if it is not forbidden by the laws of 
physics \cite{jacobson} -- might provide a way of survival for natural 
wormholes.

\acknowledgments

We are indebted to D. Monteoliva for computational help, 
and to D. F. Torres for critical reading of the draft.
LAA wishes to thank the International Centre for Theoretical Physics at 
Trieste for kind hospitality during the preparation of the manuscript.
This work has been partially supported by FOMEC (LAA), CONICET 
(MLT--HV), and  UNLP (MLT--HV--FZ). 

\begin{table}
\caption{Relation between the scalar field at the throat and its asymptotic 
value.}  
\begin{tabular}{ccc} 
$2 \omega + 3$ &   $\phi_{_\infty}$  & $\phi_{_{\rm th}}$\\
\hline
\hfill   
$ 10 \,|1 - \phi/\phi_{_\infty}^{^{\rm today}}|^{-1/2}$ &  & \\
 &  1  & 1  \\
 & .98  & .31 \\
 & .9   & .27\\
 & .8   & .22\\
\hfill
 $10^4 \,(\phi/\phi_{_\infty}^{^{\rm today}})^4$ & & \\
 &   1   &  .95\\
 &  .98  & .93\\
 & .9   & .84\\
 & .8   & .74\\
\hfill
$ 10^5 \,(\phi/\phi_{_\infty}^{^{\rm today}})^4$ & & \\
 & 1  & .98\\
 & .98  & .96\\
 & .9  & .88\\
 & .8  & .78\\
\hfill
$10\, \phi/\phi_{_\infty}^{^{\rm today}}$ & & \\
 & 1   & .06\\
 & .9  & .04
\end{tabular}
\end{table}

\begin{figure}[htbp]
\begin{center}
\epsfig{file=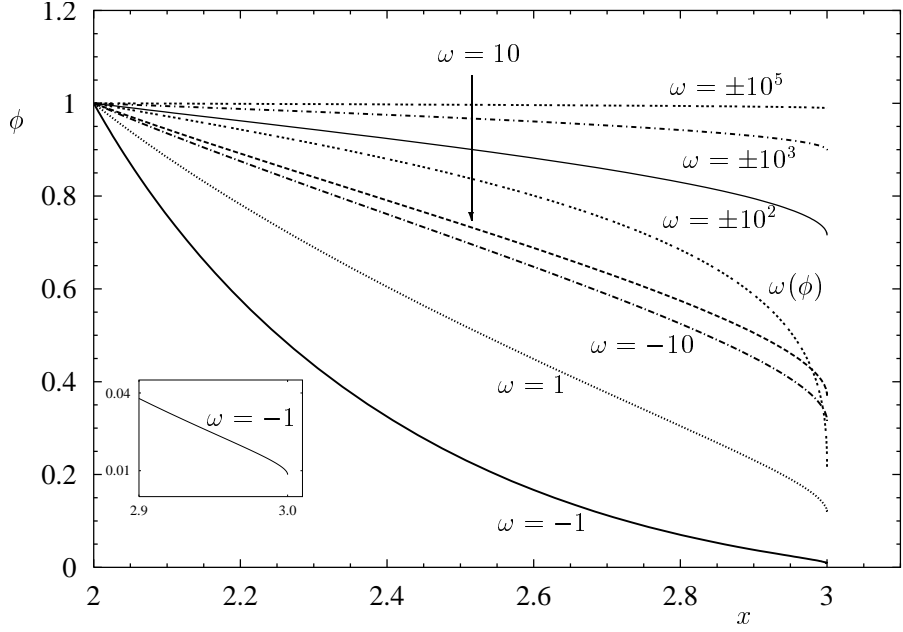,width=12cm,clip=}
\end{center}
\vspace{0.1cm}
\caption{Behavior of the scalar field as a function of the radial
coordinate for typical models of the coupling constant $\omega(\phi)$.
The embedded box details the behavior of $\phi$ when it gets close to the
wormhole throat.}
\end{figure}

\end{document}